\documentclass[twocolumn,showpacs,preprintnumbers,amsmath,amssymb]{revtex4}

\usepackage{graphicx}% Include figure files
\usepackage{dcolumn}% Align table columns on decimal point
\usepackage{bm}% bold math
\usepackage{natbib}

\begin{document}
\bibliographystyle{apsrev}

\title{Self-bound models of compact stars and recent mass-radius measurements}

\author{M. G. B. de Avellar}
\affiliation{Instituto de Astronomia, Geof\'\i sica e Ci\^encias
Atmosf\'ericas\\ Rua do
Mat\~ao 1226, 05508-900 S\~ao Paulo SP, Brazil}
\author{J. E. Horvath}
\email{foton@astro.iag.usp.br}
\affiliation{Instituto de Astronomia, Geof\'\i sica e Ci\^encias
Atmosf\'ericas\\ Rua do
Mat\~ao 1226, 05508-900 S\~ao Paulo SP, Brazil}
\author{L. Paulucci}
\email{laura.paulucci@ufabc.edu.br}
\affiliation{Universidade Federal do ABC \\
Rua Santa Ad\'elia, 166, 09210-170 Santo Andr\'e, SP, Brazil}

\date{\today}

\begin{abstract}
The exact composition of a specific class of compact stars,
historically referred to as ``neutron stars'', is still quite
unknown. Possibilities ranging from hadronic to quark degrees of
freedom, including self-bound versions of the latter have been
proposed. We specifically address the suitability of strange star
models (including pairing interactions) in this work, in the light
of new measurements available for four compact stars. The analysis
shows that these data might be explained by such an exotic
equation of state, actually selecting a small window in parameter
space, but still new precise measurements and also further
theoretical developments are needed to settle the subject.
\end{abstract}

\maketitle

\section{Introduction}

The first reasonable ideas about the composition of compact stars
stated that matter under extreme densities is composed of hadrons,
i.e., mainly neutrons with small fractions of protons and also
electrons. Further theoretical developments and modern
experimental results opened the window to other possibilities as
the existence of hyperons inside such stars due to the high
densities, and a whole new class of condensates (see, for example,
\cite{Weber} and references therein). Another possibility arisen
in the '70s is the existence of deconfined quarks inside compact
objects, either located only in the inner core of these stars or
present up to their surfaces. The latter extreme possibility was
suggested about three decades ago \cite{Itoh, Bodmer, Chin,
Terazawa, Wit}, emerging as an astrophysical realization of the
stability scenario of the so-called strange matter.

Consequences of the so-called {\it strange matter hypothesis} for
compact stars have been extensively analyzed within the simple
framework of the MIT Bag Model \cite{Farhi84, AFO} and the
Nambu-Jona-Lasinio model \cite{Alford, BenvLug98, Yin08} (see
\cite{Buballa05} for an overview on the subject). After an initial
round of perturbative considerations \cite{BailinLove,
Horvathetal91}, the possibility of non-perturbative pairing
between quarks \cite{Alford, Rapp, Alford01, RajWilc,
Rajagopal2001, German} allowed a new view on this subject due to a
great enhancement on the window of stability for strange quark
matter (having the strange quark mass, bag constant and the novel
pairing energy of the quark condensate as parameters in this
approach). This state of totally paired 3-flavor quarks is called
the color-flavor locked (CFL) strange quark matter, in which
quarks form Cooper pairs, thus lowering further the energy of the
system.

Much work has been put forward in order to have a better
understanding on the actual composition of neutron stars, both
observational and theoretical. Nevertheless, there is no absolute
conclusion for any model as yet.

We intend in this work to contribute to the understanding of how
the new and much more precise astrophysical measurements of the
mass and radius of neutron stars \cite{4U1608, EXO1745, 4U1820,
PSRJ1614} can help revealing the viability of exotic quark star
models. We analyze in particular the viability of the class of CFL
strange quark matter models. They are likely to be the most
favorable candidates for self-bound stars, in spite that more
complex models can be devised (for example intermediate phases
such as LOFF may be present \cite{LOFF1, LOFF2, Casalbuoni04}),
and they would constitute a benchmark in testing exotic compact
star composition.

In the rest of this work we quantitatively show that accurate
astrophysical measurements can already constrain important
parameters of QCD, like the Cooper pair energy gap ($\Delta$) and
the strange quark mass ($m_{s}$), and show how this is done given
a subset of the present data.

\section{Analysis}

When considering unpaired strange quark matter ($uds$ matter)
the equation of state (EoS) assumes the simplest form when
$m_{s}\rightarrow 0$:
$\epsilon=3P+4B$, being $\epsilon$, $P$ and $B$, the energy
density, pressure and bag constant, respectively. The
non-zero strange quark mass has the effect of increasing the
energy of the system \cite{Farhi84}. On the other hand,
if the system presents pairing interactions between quarks it has the
net effect of lowering the energy of the system, making it more
stable.

There have been different approaches to analyze this effect. Here,
in order to set a framework for the class of models studied, we
adopt the MIT bag model thermodynamic potential of CFL strange
quark matter \cite{Rajagopal2001, Alford01} in bulk to order
$\Delta^2$, being $\Delta$ the pairing energy gap for the phase,
given by

\begin{equation}
\Omega_{CFL}=\Omega_{free}-\frac{3}{\pi^2}\Delta^2\mu^2+B \label{CFL_EOS}
\end{equation}

In this expression, the CFL state is taken as a model composed of
quark matter without pairing plus a term corresponding to the
energy of the diquark condensate. Being so, the pairing energy gap
is taken as a free parameter, instead of being defined in a
self-consistent manner (see, for example, \cite{Paulucci2011} for
further discussions and more consistent models). The thermodynamic
potential for the unpaired state is then

\begin{equation}
\Omega_{free}=\sum_{i=u,d,s}\frac{1}{4\pi^2}\Big[\mu_i\nu\Big(\mu_i^2-\frac{5}{2}m_i^2\Big)+
\frac{3}{2}m_i^4\log\Big(\frac{\mu_i+\nu}{m_i}\Big)\Big]
\end{equation}
\\
with $\mu=(\mu_u+\mu_d+\mu_s)/3$ and the common Fermi momentum,
$\nu=2\mu-\sqrt{\mu^2+m_s^2/2}$.

The EoS is finally obtained by relating the energy
density $\epsilon$ to the pressure $P$ as follows

\begin{equation}
\epsilon=3\mu n_B - P\label{eos}
\end{equation}
\\
with the particle density given by $n_B=(\nu^3+2\Delta^2\mu)/ \pi^2$,
and the pressure $P=-\Omega_{CFL}$.

Due to the fact that $m_{s}\neq 0$, the EoS have to be computed
numerically. However, the CFL EoS shows a remarkably linear
behavior \cite{German}. Because of this, much work has been done in
order to construct a simple EoS that includes the effects of pairing and
a heavy strange quark, keeping the linear behavior to a certain
extent. We start our analysis, then, by postulating a linear EoS
for CFL matter.

We have parameterized the equation of state for CFL matter as
\begin{equation}
P=\alpha(c^2\rho-4B_{eff}), \label{Parametrizada}
\end{equation}
\\
where $\alpha$ is a parameter
related with the stiffness of the EoS and $B_{eff}$ is certainly related
to the bag constant $B$ and gap parameter $\Delta$.
Within this approximation,
it is easy to show that the Tolman-Oppenheimer-Volkoff (TOV) and
the EoS equations can be written in dimensionless forms:

\begin{equation}
p^{'}_{\star}=-\frac{m_{\star}\rho_{\star}}{x^{2}}
\Bigg(1+\frac{p_{\star}}{\rho_{\star}}\Bigg)\Bigg(1+\frac{4\pi x^{3}p_{\star}}{m_{\star}}\Bigg)
\Bigg(1-2\frac{m_{\star}}{x}\Bigg)^{-1}
\end{equation}

\begin{equation}
m^{'}_{\star}=4\pi x^{2}\rho_{\star}
\end{equation}

\begin{equation}
p_{\star}=\alpha (\rho_{\star}-1)
\end{equation}
\\
where the starred quantities are the dimensionless mass, pressure
and mass density; $x$ is the dimensionless radius and
$\alpha\equiv\frac{1}{3+\beta}$ is an effective parameter which
depends on the quantities $\Delta$ and $m_{s}$ in a complicated
but very mild and continuous way.

Our main goal here was to devise an easy way to display the
location ({\it locus}) of the maximum mass and its correspondent
radius, e. g., fitting a curve $f=f(x,\alpha)$. This {\it locus}
strongly depends, in our parameterization, on $\alpha$. The higher
its value, the higher the maximum mass. Because of this, we needed
to construct several mass-radius relationships varying $\alpha$
within an acceptable range, in this case, from 0.20 to 0.36. To
comply with observations, we needed to reach at least
$2M_{\odot}$. We then fitted an expression for the maximum mass
($m_{\star,max}$) of each sequence and their correspondent radius
($x_{corr}$) dependent on $\alpha$, yielding the result

\begin{eqnarray}
m_{\star,max}=-0.163\alpha^{2} + 0.201\alpha + 0.00286\label{MRadmEq}\\
x_{corr}=-0.262\alpha^{2} + 0.329\alpha +0.111.
\end{eqnarray}

Restoring the units, we finally find the relation

\begin{equation}\label{MRdim}
M_{max}=R_{corr}\frac{c^{2}}{G}\frac{m_{max}}{x_{corr}}.
\end{equation}
given that

$$
r=x\sqrt{\frac{c^{4}}{4B_{eff}G}}\hspace{0.4cm}m=m_{\star}\sqrt{\frac{c^{8}}{4B_{eff}G^{3}}}.
$$

Here is the point where the dependence on $B_{eff}$ appears.
Notice that the radius and the mass individually depend on
$B_{eff}$ but the relation between maximum mass and its
correspondent radius (when the physical units are restored) does
{\it not}: it depends only on the value of $\alpha$. This feature
is already present in Witten's original paper and seems to be
generic for any linear equation of state.

From the effective parameters found above, we have analyzed the
agreement of the mass-radius relation obtained for different sets
of parameters with data for the following stars: 4U 1608-522
\cite{4U1608}, EXO 1745-248 \cite{EXO1745}, 4U 1820-30
\cite{4U1820} and PSR J1614-2230 \cite{PSRJ1614}. The results are
presented in the following section.

\section{Results}

\begin{figure*}
\includegraphics[width=0.46\textwidth]{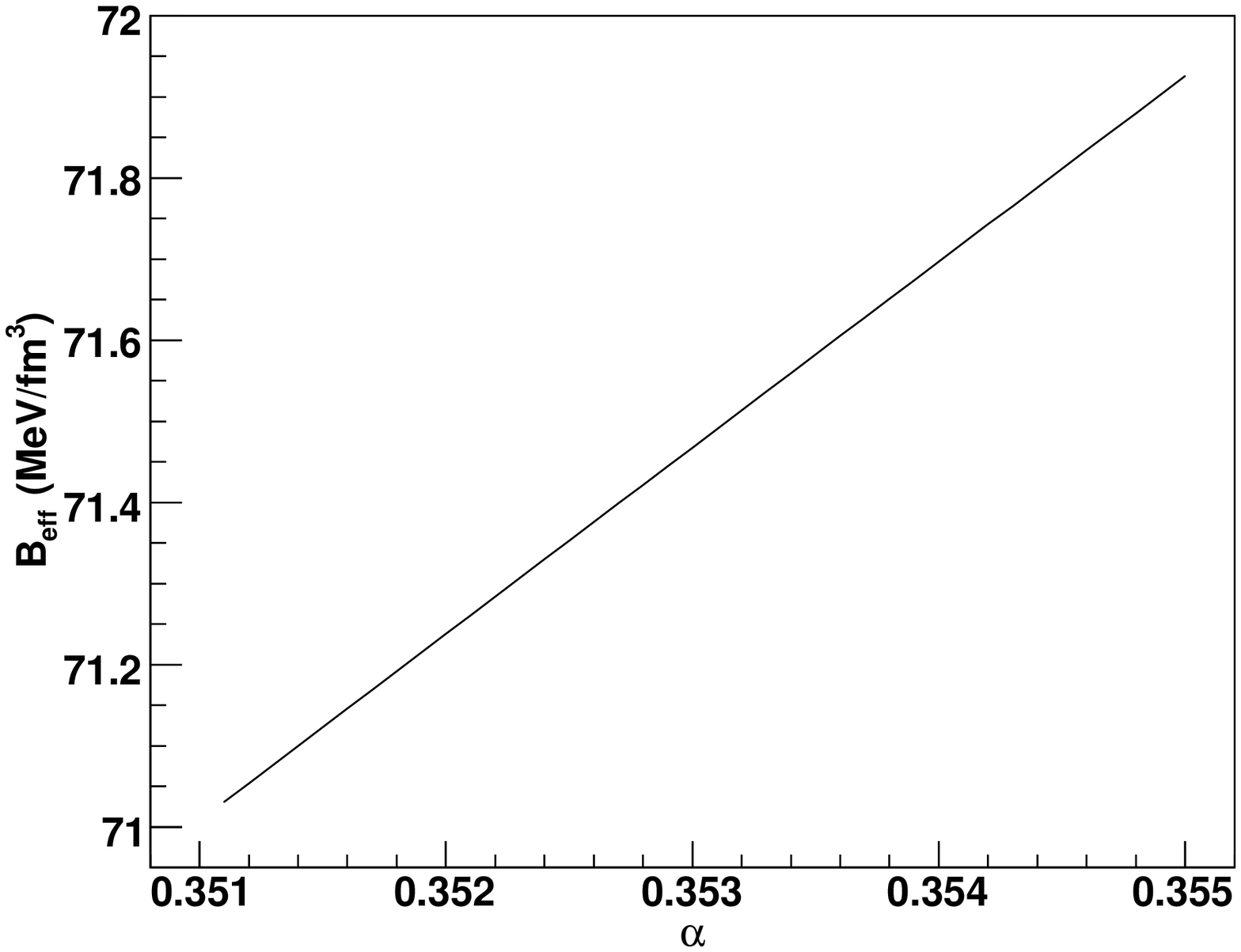}
\includegraphics[width=0.49\textwidth]{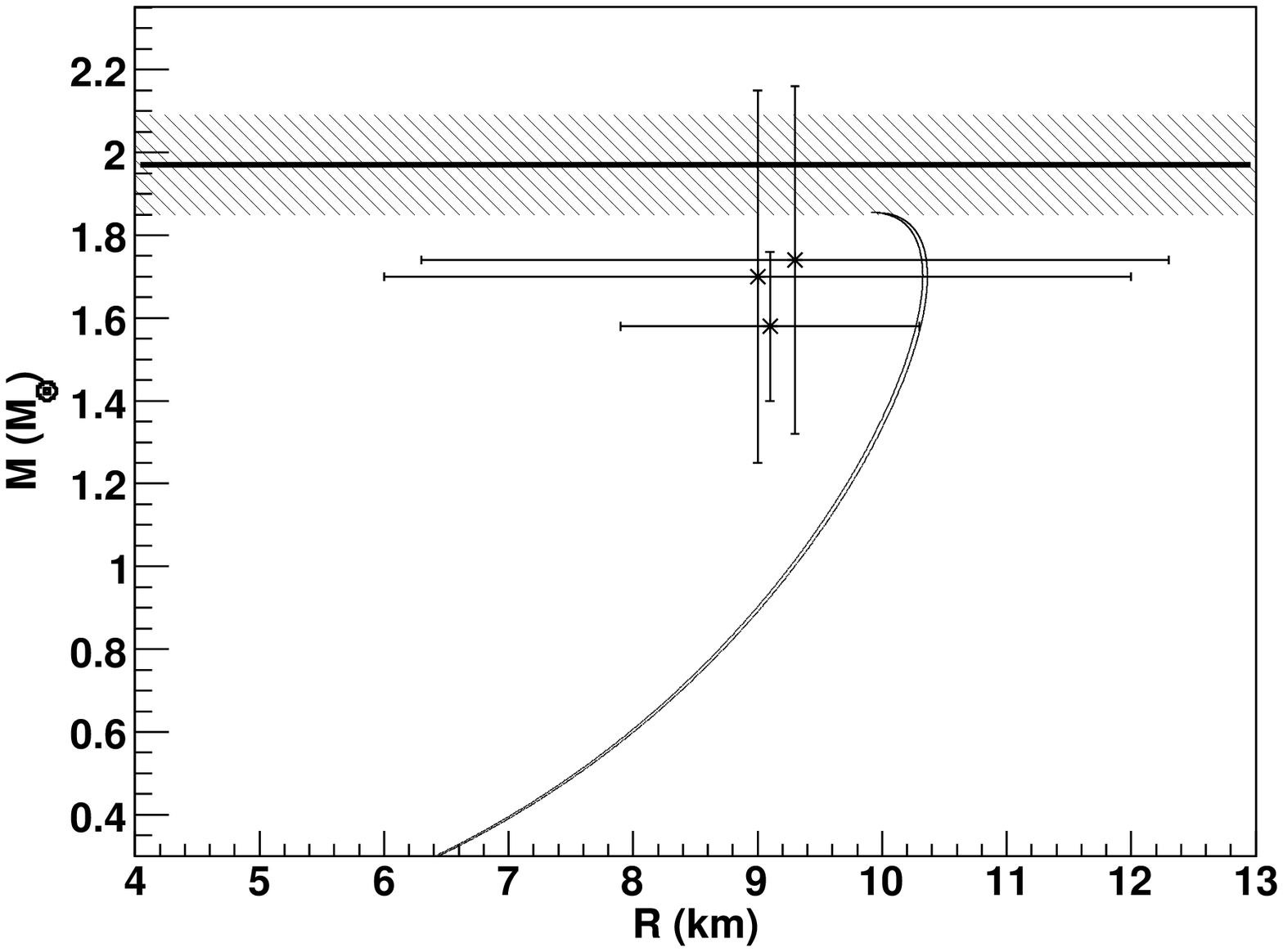}
\caption{On the left panel, effective parameters allowing
compatibility within 3$\sigma$ with recent data (taking into
account the observed masses and radii and their respective error bars). On
the right, mass-radius measurements are shown as crosses with
3$\sigma$ error bars. Data from PSR J1614-2230 is shown as a
shaded region, since there is only information on the mass for
this object. The two mass-radius relations shown correspond to the
lower and upper limits of $\alpha$ and $B_{eff}$ that can explain
the data (shown on the right).}\label{MRadm}
\end{figure*}

The value of $\alpha$ as a function of $B_{eff}$ can be
constrained by using the observed mass of PSR J1614-2230,
considering that the maximum mass provided by a specific EoS
should at least equal the mass of this heavy star. This can be
obtained from eq. (\ref{MRadmEq})-(\ref{MRdim}) or by varying the
values of $\alpha$ and $B_{eff}$ on Eq. (\ref{Parametrizada}) and
constructing the corresponding mass-radius sequences. Within this
previously determined region and having generated the mass-radius
relations for the effective equation of state using these
constrained values, we compared them with the other three stars
within a 3$\sigma$ demanded precision. The effective parameters,
i.e. all the pairs ($\alpha,B_{eff}$), that would be suitable for
explaining all four data at the same time, i.e. the observed
masses and radii and their respective error bars, are those corresponding to
the line shown on the left panel of Fig. \ref{MRadm}.
The initial and final points of this line were used to generate 
the mass-radius relations shown on the right panel of the same Fig.

We finally find a very narrow window for $\alpha$ and $B_{eff}$
that can provide compatibility with all four measurements
together. As a consequence, the mass-radius relations shown on the
right panel of Fig. \ref{MRadm} are almost completely overlapping.

We would like now to relate the effective parameters determined
through this analysis with the free parameters determining the CFL
matter state. Unfortunately, although the EoS presents a linear
behavior, the linearization of Eq.(\ref{CFL_EOS}) is not possible
analytically, unless some major approximations are made. Without
the latter, one cannot directly relate $\alpha$ and $B_{eff}$ to
$\Delta$, $m_s$ and $B$. Previous attempts to find such a relation
include the work of Alford et al. \cite{AlfordReddy}, where a form
of parameterized EoS was proposed, in which QCD corrections to the
pressure, represented by a parameter $c\approx 0.3$, have been
added. This parameterization relied on a power series expansion in
$\mu$ for the thermodynamic potential of the quark phase, reading

\begin{equation}
\Omega_{QM}=-\frac{3}{4\pi^2}a_4\mu^4+\frac{3}{4\pi^2}a_2\mu^2+B_{eff}
\end{equation}
\\
with $a_4=1-c$ and $a_2$ as free parameters. In their approach,
the strange quark mass and gap parameter affect $a_2$ alone and
$B_{eff}$ is determined by the density at which the transition
between quark and nuclear matter phases occur, as they were
interested in addressing hybrid stars. Although they have been
able to write $a_2=a_2(m_s,\Delta)$, and one can in fact for CFL
quark matter in equilibrium (within this framework) separate the
thermodynamic potential expression to order $m_s^4$ with only
three terms proportional to $\mu^4$, $\mu^2$ and $\mu^0$, the
chemical potential itself has a nonlinear and nontrivial
dependence on all other parameters, as shown in details in
\cite{German}. We have chosen {\it not} to make this approximation
here, so that we are not able to write $\alpha$ and $B_{eff}$ as
function of $m_s$, $B$ and $\Delta$ in analytical closed form.

Since it is not possible to obtain the expression relating the
parameterized EoS to the CFL one, in order to verify the
effectiveness of our own linearization, we have taken the full
equation of state for CFL strange quark matter, as written in Eq.
(\ref{CFL_EOS})-(\ref{eos}), and analyzed the agreement of the
mass-radius relation obtained for the same set of stars as before.
We have covered acceptable ranges for the free parameters: the bag
constant $B > 57$ MeV/fm$^3$, the superconductor gap
$0\leq\Delta\lesssim 50$ MeV and the strange quark mass $m_s\geq
100$ MeV. Recent measurements give a very precise value for the
$s$ quark mass at the energy scale of $2$ GeV: $92.4\pm 1.5$ MeV
\cite{HPQCD}. At a smaller energy scale this value should be even
higher (see \cite{PDG}, Particle Properties - quark masses, and
references therein) so we decided to adopt the value of 100 MeV as
a lower boundary.

\begin{figure}
\includegraphics[width=0.49\textwidth]{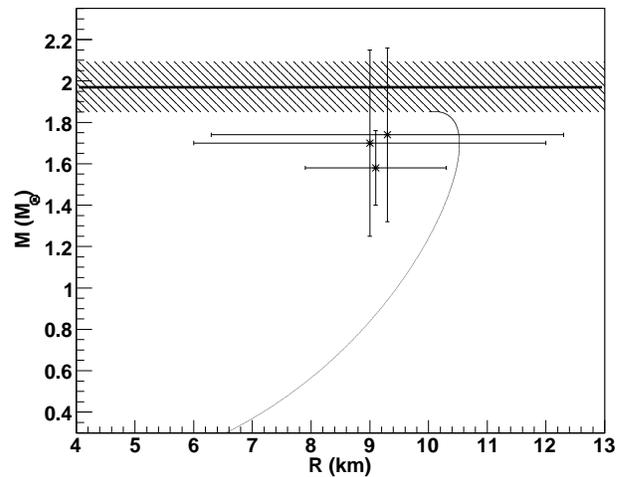}
\caption{\label{CFL}Best agreement for the mass-radius relation
with recent measurements (data represented as in Fig. \ref{MRadm})
considering the CFL equation of state.
Values for the parameters are $B\sim 69$ MeV/fm$^3$,
$\Delta \sim 50$ MeV, and $m_s\sim 120$ MeV.}
\end{figure}

After solving the TOV equations for several sets of ($m_s$, $B$,
$\Delta$) and then comparing the values of mass and radius with
the measurements under consideration, the best agreement found is
the one giving the mass-radius relation shown in Fig. \ref{CFL}.
The ``optimal'' values of the parameters found are $B\sim 69$
MeV/fm$^3$, $\Delta \sim 50$ MeV, and $m_s\sim 120$ MeV (well
within the window of stability for CFL strange quark matter, thus
justifying the quark star scenario). Notice, however, that for 4U
1820-30 the mass-radius curve is in agreement within $\sim
3.4\sigma$ only. It should also be noticed, nevertheless, that one
cannot state that in this case the $3\sigma$ boundary is the one
corresponding to 99.9\% of compatibility, since there are very few
degrees of freedom involved in the analysis. Therefore, the
agreement can be considered as acceptable.

We see that, although the equation of state for CFL strange quark
matter clearly presents a linear behavior in the pressure versus
energy density dependency, when a parameterization of the type
$P=\alpha(c^{2}\rho-4B_{eff})$ is made, the coefficients $\alpha$
and $B_{eff}$ cannot be treated as independent from each other.
Since we are working with a ``toy model'', in which one considers
a system without any pairing plus a term corresponding to the
condensate energy, $\Delta$ surely adds up to the bag constant but
has no influence on the ``unpaired state''. However, the role of
the strange quark mass is more subtle. In the analysis discussed
in \cite{AlfordReddy}, the authors show that $B_{eff}$, besides
being related to the vacuum pressure (and, in the CFL phase, also
to the gap parameter), incorporates an additional term which is
proportional to $m_s^4$ when the power series expansion is made.
So even in a simplified scheme, we see that $B_{eff}$ is not
dependent on the bag constant alone but should also incorporate
additional terms which are important for determining the stiffness
of the EoS.

In this way, even having found values for the parameters in the
linearization that could fit the data within a $\sim 3\sigma$
region, when a full equation of state is employed, one cannot
reach such a good agreement. This kind of problem is also evident
in the oldest expression employed by Benvenuto and Horvath
\cite{BH89} $P=\frac{1}{3+a}[c^{2}\rho -(4+b)B]$, the complicated
dependence of the parameters on $m_{s}$, $B$ and $\Delta$ (and
hence their correlation) affects the accuracy of the $M_{max}-R$
{\it locus} and blurs to some extent an accurate determination of
the parameter space.

Another general aspect worth of noticing is the very soft raise of
the mass-radius curve for paired strange quark matter. It could be
attributed to the simplicity of the model, not considering
interactions between quarks other than pairing and one-gluon
exchange (see \cite{Farhi84} for a discussion on the scaling of
the bag parameter with $\alpha_c \neq 0$). A steeper raise of the
mass-radius relation could indeed provide a better agreement with
the observed data and perhaps an increase in the maximum allowed
mass. This could be an important issue since it would be
impossible to reach masses as high as 2$M_{\odot}$, for example,
with a single quark matter EoS that would also need to explain the
values for measured masses and radii for the stars 4U 1608-522,
EXO 1745-248, and 4U 1820-30, if they happen ultimately to be
accurate and correct.

\begin{figure*}
\includegraphics[width=0.49\textwidth]{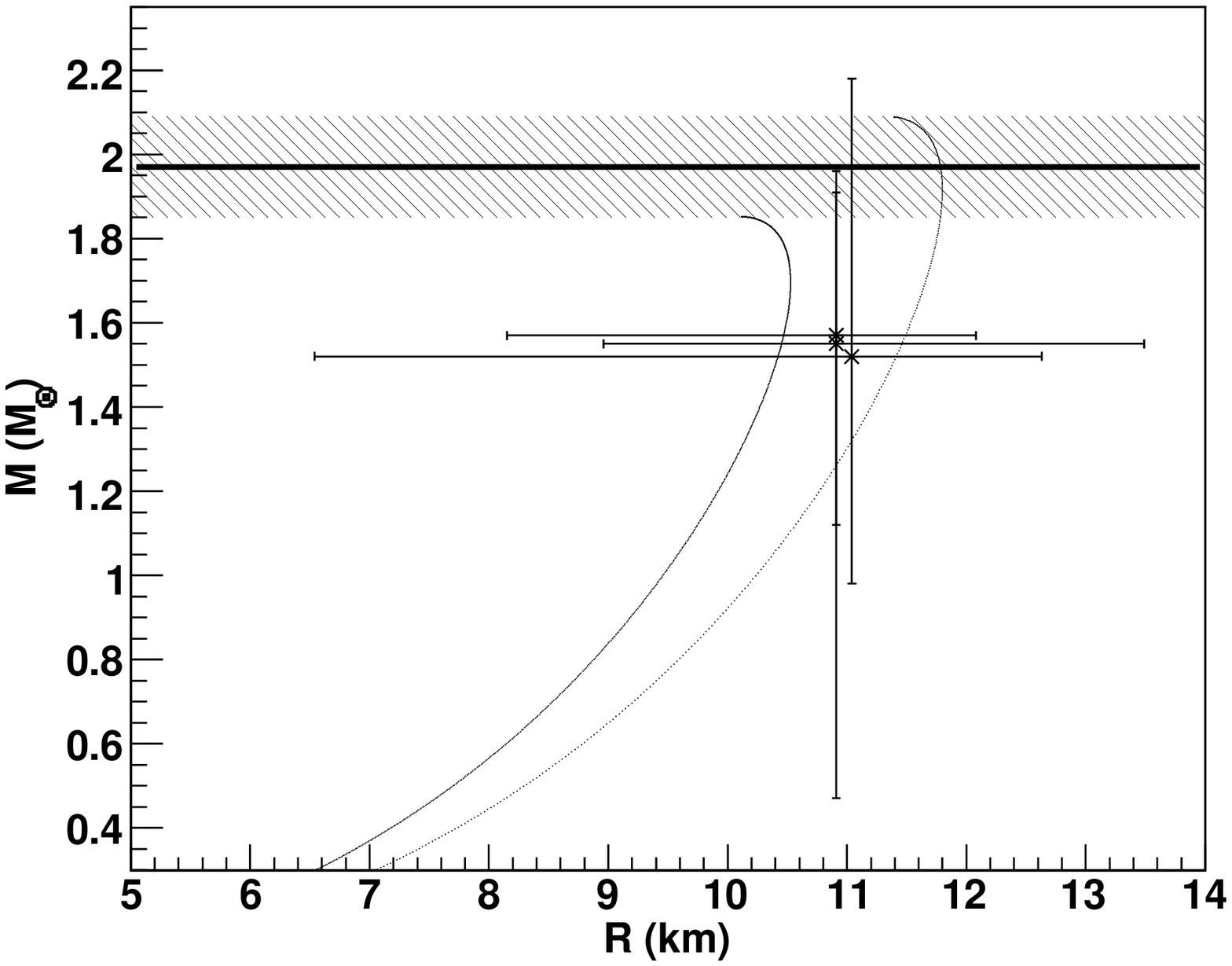}
\includegraphics[width=0.49\textwidth]{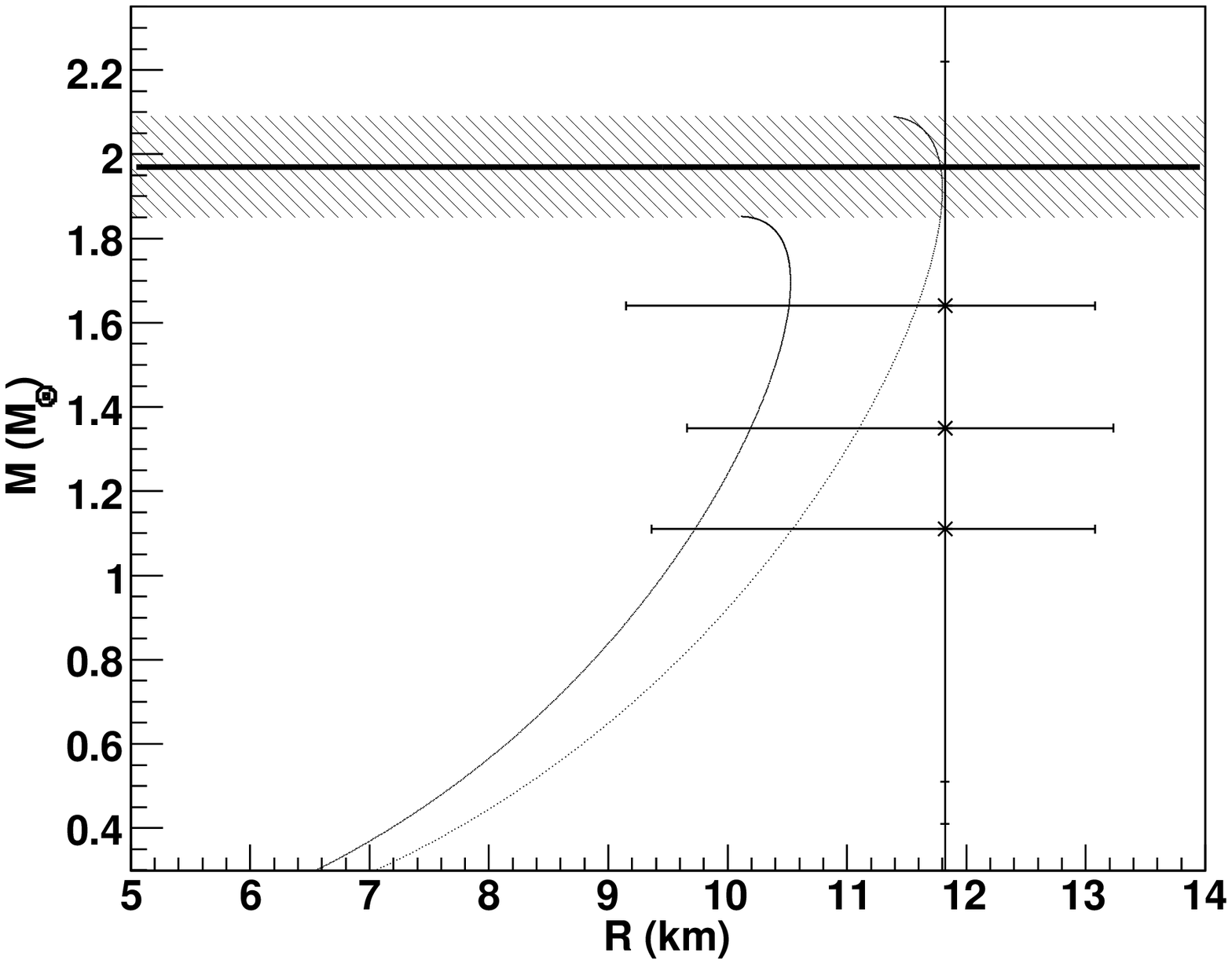}
\caption{\label{Lattimer1} A second evaluation of the
reconstructed data as given in \cite{Lattimer}. On the left side,
mass and radius of 4U 1608-522, EXO 1745-248 and 4U 1820-30 and
mass range for PSR J1614-2230 considering $R=R_{ph}$. On the
right, the same for $R>>R_{ph}$ (see text for further details).
Two examples of mass-radius relations for CFL stars are shown for the sets of 
parameters ($B$, $\Delta$, $m_s$) given by (69 MeV/fm$^3$, 50 MeV, 120 MeV)
and (58 MeV/fm$^3$, 62 MeV, 110 MeV) to highlight the fact that, 
in this case, the mass-radius relation is constrained
solely by the massive pulsar.}
\end{figure*}

There have been claims, however, that the values of masses and
radii for 4U 1608-522, EXO 1745-248 and 4U 1820-30 have been
mistakenly derived \cite{Lattimer}. The computational method used
in the previous analysis was criticized pointing out possible
systematic errors which substantially reduce the uncertainties in
the masses and radii values for these systems. Considering in turn
the determinations given by Steiner, Lattimer and Brown \cite{Lattimer},
we have
repeated the previous calculations to verify whether the CFL equation
of state is compatible with the reconstruction of the data
presented by them. The results of our analysis for this case are
very different, though. We see that the very narrow parameter
space previously found now becomes an extremely wide region of
possibilities, as depicted in Fig \ref{Lattimer1}.

The error bars for these systems, according to Steiner, Lattimer
and Brown, are much bigger than previously
considered and all three stars are clustered with close values for
their masses and radii ($M\sim 1.5 M_{\odot}$ and $R\sim 11$ km
with more visible differences for the mass when the photospheric
radius for photon emission is considered to be much larger than
the stellar radius, $R_{ph}>>R$, the color correction $f_{c}$ is a
key ingredient for this behavior). This leads now to a large set
of parameters $m_{s}$, $B$ and $\Delta$ that could in principle
explain the data, basically having the mass of PSR J1614-2230 as
the main quantity providing the constrain for the equation of
state. In other words, one can find several sets of (B, $\Delta$,
$m_s$) that satisfy the necessary conditions within the values
needed in this analysis. The general behavior is to have a lower
maximum mass with an increase in the strange quark mass and bag
constant when the values of the other two variables are held
fixed. As the gap parameter increases, the the maximum allowed
mass also increases.

\section{Conclusions}

Precise measurements for masses and radii of compact stars are
very important for constraining their composition. However,
besides being precise, it is necessary that they also be accurate
for a clear comparison of the mass-radius relation coming from
very different compositions with data.

Given the present controversy of the radius determination quoted
in the paper, this appears to be the main issue to be resolved for
an advance of the conclusions of studies like ours. Meantime we
can only conclude that only with small error bars for both mass
and radii determinations (as in the analysis of \cite{4U1608,
EXO1745, 4U1820}) the parameters of the CFL equation of state can
be effectively constrained. Otherwise, several sets of parameters
defining the CFL EoS can be used to match the data. In this case,
more advanced detection techniques and physical analysis are
required to constrain the radius $R_{ph}$ in X-ray burst models.

Since one should {\it not} treat $\alpha$ and $B_{eff}$ as
independent quantities in a parameterized EoS for CFL strange
quark matter like the one proposed here, an analysis performed
with a parameterization with correlated coefficients should be
explored. We intend to address this subject in a future work.

An additional interesting report is the very low mass of at least
one neutron star mass observed ($M=0.87\pm 0.07 M_{\odot}$ for
4U 1538-52) \cite{Rawls}, actually not well explained by ordinary
stellar evolution arguments. This ``low-end'' mass extreme is also
important for the quest of exotic compositions in their interiors.
In fact, confirming such a measurement would imply a radius of $R
\leq 9$ km within the self-bound hypothesis, while the measurement
of, say, $R \geq 11$ km would be totally inconsistent with it,
pointing to a normal composition. We are at the edge of finally
addressing the fundamental composition of compact stars, at least
on a first level that leaves a handful of possibilities left out
of a huge number of proposals.

\begin{acknowledgments}
We wish to acknowledge the support of the Funda\c c\~ao de Amparo
\`a Pesquisa do Estado de S\~ao Paulo and the CNPq Agency and CAPES
Agency (Brazil). We would also like to thank Dr. Franciole Marinho for
very useful discussions on statistical analysis.
\end{acknowledgments}

\bibliography{AHP}

\begin{thebibliography}{33}
\expandafter\ifx\csname natexlab\endcsname\relax\def\natexlab#1{#1}\fi
\expandafter\ifx\csname bibnamefont\endcsname\relax
  \def\bibnamefont#1{#1}\fi
\expandafter\ifx\csname bibfnamefont\endcsname\relax
  \def\bibfnamefont#1{#1}\fi
\expandafter\ifx\csname citenamefont\endcsname\relax
  \def\citenamefont#1{#1}\fi
\expandafter\ifx\csname url\endcsname\relax
  \def\url#1{\texttt{#1}}\fi
\expandafter\ifx\csname urlprefix\endcsname\relax\def\urlprefix{URL }\fi
\providecommand{\bibinfo}[2]{#2}
\providecommand{\eprint}[2][]{\url{#2}}

\bibitem[{\citenamefont{Weber}(2005)}]{Weber}
\bibinfo{author}{\bibfnamefont{F.}~\bibnamefont{Weber}},
  \bibinfo{journal}{Prog. Part. Nucl. Phys.} \textbf{\bibinfo{volume}{54}},
  \bibinfo{pages}{193} (\bibinfo{year}{2005}).

\bibitem[{\citenamefont{Itoh}(1970)}]{Itoh}
\bibinfo{author}{\bibfnamefont{N.}~\bibnamefont{Itoh}},
  \bibinfo{journal}{Progr. Theor. Phys.} \textbf{\bibinfo{volume}{44}},
  \bibinfo{pages}{291} (\bibinfo{year}{1970}).

\bibitem[{\citenamefont{Bodmer}(1971)}]{Bodmer}
\bibinfo{author}{\bibfnamefont{A.}~\bibnamefont{Bodmer}},
  \bibinfo{journal}{Phys. Rev. D} \textbf{\bibinfo{volume}{4}},
  \bibinfo{pages}{1601} (\bibinfo{year}{1971}).

\bibitem[{\citenamefont{Chin and Kerman}(1979)}]{Chin}
\bibinfo{author}{\bibfnamefont{S.~A.} \bibnamefont{Chin}} \bibnamefont{and}
  \bibinfo{author}{\bibfnamefont{A.~K.} \bibnamefont{Kerman}},
  \bibinfo{journal}{Phys. Rev. Lett.} \textbf{\bibinfo{volume}{43}},
  \bibinfo{pages}{1292} (\bibinfo{year}{1979}).

\bibitem[{\citenamefont{Terazawa}(1979)}]{Terazawa}
\bibinfo{author}{\bibfnamefont{H.}~\bibnamefont{Terazawa}},
  \bibinfo{journal}{Tokyo U. Report.} pp. \bibinfo{pages}{INS--336}
  (\bibinfo{year}{1979}).

\bibitem[{\citenamefont{Witten}(1984)}]{Wit}
\bibinfo{author}{\bibfnamefont{E.}~\bibnamefont{Witten}},
  \bibinfo{journal}{Phys. Rev. D} \textbf{\bibinfo{volume}{30}},
  \bibinfo{pages}{272} (\bibinfo{year}{1984}).

\bibitem[{\citenamefont{Farhi and Jaffe}(1984)}]{Farhi84}
\bibinfo{author}{\bibfnamefont{E.}~\bibnamefont{Farhi}} \bibnamefont{and}
  \bibinfo{author}{\bibfnamefont{R.~L.} \bibnamefont{Jaffe}},
  \bibinfo{journal}{Phys. Rev. D} \textbf{\bibinfo{volume}{30}},
  \bibinfo{pages}{2379} (\bibinfo{year}{1984}).

\bibitem[{\citenamefont{Alcock et~al.}(1986)\citenamefont{Alcock, Farhi, and
  Olinto}}]{AFO}
\bibinfo{author}{\bibfnamefont{C.}~\bibnamefont{Alcock}},
  \bibinfo{author}{\bibfnamefont{E.}~\bibnamefont{Farhi}}, \bibnamefont{and}
  \bibinfo{author}{\bibfnamefont{A.~V.} \bibnamefont{Olinto}},
  \bibinfo{journal}{Astrophys. Journal} \textbf{\bibinfo{volume}{310}},
  \bibinfo{pages}{261} (\bibinfo{year}{1986}).

\bibitem[{\citenamefont{Alford et~al.}(1999)\citenamefont{Alford, Rajagopal,
  and Wilczek}}]{Alford}
\bibinfo{author}{\bibfnamefont{M.}~\bibnamefont{Alford}},
  \bibinfo{author}{\bibfnamefont{K.}~\bibnamefont{Rajagopal}},
  \bibnamefont{and} \bibinfo{author}{\bibfnamefont{F.}~\bibnamefont{Wilczek}},
  \bibinfo{journal}{Nuc. Phys.} \textbf{\bibinfo{volume}{B537}},
  \bibinfo{pages}{433} (\bibinfo{year}{1999}).

\bibitem[{\citenamefont{Benvenuto and Lugones}(1998)}]{BenvLug98}
\bibinfo{author}{\bibfnamefont{O.~G.} \bibnamefont{Benvenuto}}
  \bibnamefont{and} \bibinfo{author}{\bibfnamefont{G.}~\bibnamefont{Lugones}},
  \bibinfo{journal}{Int. J. Mod. Phys.} \textbf{\bibinfo{volume}{7}},
  \bibinfo{pages}{29} (\bibinfo{year}{1998}).

\bibitem[{\citenamefont{Yin and Su}(2008)}]{Yin08}
\bibinfo{author}{\bibfnamefont{S.}~\bibnamefont{Yin}} \bibnamefont{and}
  \bibinfo{author}{\bibfnamefont{R.-K.} \bibnamefont{Su}},
  \bibinfo{journal}{Phys. Rev. C} \textbf{\bibinfo{volume}{77}},
  \bibinfo{pages}{055204} (\bibinfo{year}{2008}).

\bibitem[{\citenamefont{Buballa}(2005)}]{Buballa05}
\bibinfo{author}{\bibfnamefont{M.}~\bibnamefont{Buballa}},
  \bibinfo{journal}{Phys. Rept.} \textbf{\bibinfo{volume}{407}},
  \bibinfo{pages}{205} (\bibinfo{year}{2005}).

\bibitem[{\citenamefont{Bailin and Love}(1984)}]{BailinLove}
\bibinfo{author}{\bibfnamefont{D.}~\bibnamefont{Bailin}} \bibnamefont{and}
  \bibinfo{author}{\bibfnamefont{A.}~\bibnamefont{Love}},
  \bibinfo{journal}{Phys. Rep.} \textbf{\bibinfo{volume}{107}},
  \bibinfo{pages}{325} (\bibinfo{year}{1984}), \bibinfo{note}{and references
  therein.}

\bibitem[{\citenamefont{Horvath et~al.}(1991)\citenamefont{Horvath, Benvenuto,
  and Vucetich}}]{Horvathetal91}
\bibinfo{author}{\bibfnamefont{J.~E.} \bibnamefont{Horvath}},
  \bibinfo{author}{\bibfnamefont{O.~G.} \bibnamefont{Benvenuto}},
  \bibnamefont{and} \bibinfo{author}{\bibfnamefont{H.}~\bibnamefont{Vucetich}},
  \bibinfo{journal}{Phys. Rev. D} \textbf{\bibinfo{volume}{44}},
  \bibinfo{pages}{3797} (\bibinfo{year}{1991}).

\bibitem[{\citenamefont{Rapp et~al.}(2000)\citenamefont{Rapp, Schaefer,
  Shuryak, and Velkovsky}}]{Rapp}
\bibinfo{author}{\bibfnamefont{R.}~\bibnamefont{Rapp}},
  \bibinfo{author}{\bibfnamefont{T.}~\bibnamefont{Schaefer}},
  \bibinfo{author}{\bibfnamefont{E.~V.} \bibnamefont{Shuryak}},
  \bibnamefont{and}
  \bibinfo{author}{\bibfnamefont{M.}~\bibnamefont{Velkovsky}},
  \bibinfo{journal}{Ann. Phys. (NY)} \textbf{\bibinfo{volume}{280}},
  \bibinfo{pages}{35} (\bibinfo{year}{2000}).

\bibitem[{\citenamefont{Alford et~al.}(2001{\natexlab{a}})\citenamefont{Alford,
  Rajagopal, Reddy, and Wilczek}}]{Alford01}
\bibinfo{author}{\bibfnamefont{M.}~\bibnamefont{Alford}},
  \bibinfo{author}{\bibfnamefont{K.}~\bibnamefont{Rajagopal}},
  \bibinfo{author}{\bibfnamefont{S.}~\bibnamefont{Reddy}}, \bibnamefont{and}
  \bibinfo{author}{\bibfnamefont{F.}~\bibnamefont{Wilczek}},
  \bibinfo{journal}{Phys. Rev. D} \textbf{\bibinfo{volume}{64}},
  \bibinfo{pages}{074017} (\bibinfo{year}{2001}{\natexlab{a}}).

\bibitem[{\citenamefont{Rajagopal and Wilczeck}()}]{RajWilc}
\bibinfo{author}{\bibfnamefont{K.}~\bibnamefont{Rajagopal}} \bibnamefont{and}
  \bibinfo{author}{\bibfnamefont{F.}~\bibnamefont{Wilczeck}},
  \bibinfo{howpublished}{arXiv:hep-ph/0011333, and references therein}.

\bibitem[{\citenamefont{Rajagopal and Wilczek}(2001)}]{Rajagopal2001}
\bibinfo{author}{\bibfnamefont{K.}~\bibnamefont{Rajagopal}} \bibnamefont{and}
  \bibinfo{author}{\bibfnamefont{F.}~\bibnamefont{Wilczek}},
  \bibinfo{journal}{Phys. Rev. Lett.} \textbf{\bibinfo{volume}{86}},
  \bibinfo{pages}{3492} (\bibinfo{year}{2001}).

\bibitem[{\citenamefont{Lugones and Horvath}(2002)}]{German}
\bibinfo{author}{\bibfnamefont{G.}~\bibnamefont{Lugones}} \bibnamefont{and}
  \bibinfo{author}{\bibfnamefont{J.~E.} \bibnamefont{Horvath}},
  \bibinfo{journal}{Phys. Rev. D} \textbf{\bibinfo{volume}{66}},
  \bibinfo{pages}{074017} (\bibinfo{year}{2002}).

\bibitem[{\citenamefont{Guver et~al.}(2010{\natexlab{a}})\citenamefont{Guver,
  Ozel, Cabrera-Lavres, and Wroblewski}}]{4U1608}
\bibinfo{author}{\bibfnamefont{T.}~\bibnamefont{Guver}},
  \bibinfo{author}{\bibfnamefont{F.}~\bibnamefont{Ozel}},
  \bibinfo{author}{\bibfnamefont{A.}~\bibnamefont{Cabrera-Lavres}},
  \bibnamefont{and}
  \bibinfo{author}{\bibfnamefont{P.}~\bibnamefont{Wroblewski}},
  \bibinfo{journal}{Astrophys. J.} \textbf{\bibinfo{volume}{712}},
  \bibinfo{pages}{964} (\bibinfo{year}{2010}{\natexlab{a}}),
  \bibinfo{note}{arXiv:0811.3979v4 [astro-ph]}.

\bibitem[{\citenamefont{Ozel et~al.}(2009)\citenamefont{Ozel, Guver, and
  Psaltis}}]{EXO1745}
\bibinfo{author}{\bibfnamefont{F.}~\bibnamefont{Ozel}},
  \bibinfo{author}{\bibfnamefont{T.}~\bibnamefont{Guver}}, \bibnamefont{and}
  \bibinfo{author}{\bibfnamefont{D.}~\bibnamefont{Psaltis}},
  \bibinfo{journal}{Astrophys. J.} \textbf{\bibinfo{volume}{693}},
  \bibinfo{pages}{1775} (\bibinfo{year}{2009}).

\bibitem[{\citenamefont{Guver et~al.}(2010{\natexlab{b}})\citenamefont{Guver,
  Wroblewski, Camarota, and Ozel}}]{4U1820}
\bibinfo{author}{\bibfnamefont{T.}~\bibnamefont{Guver}},
  \bibinfo{author}{\bibfnamefont{P.}~\bibnamefont{Wroblewski}},
  \bibinfo{author}{\bibfnamefont{L.}~\bibnamefont{Camarota}}, \bibnamefont{and}
  \bibinfo{author}{\bibfnamefont{F.}~\bibnamefont{Ozel}},
  \bibinfo{journal}{Astrophys. J.} \textbf{\bibinfo{volume}{719}},
  \bibinfo{pages}{1807} (\bibinfo{year}{2010}{\natexlab{b}}).

\bibitem[{\citenamefont{Demorest et~al.}(2010)\citenamefont{Demorest, Pennucci,
  Ransom, Roberts, and Hessels}}]{PSRJ1614}
\bibinfo{author}{\bibfnamefont{P.~B.} \bibnamefont{Demorest}},
  \bibinfo{author}{\bibfnamefont{T.}~\bibnamefont{Pennucci}},
  \bibinfo{author}{\bibfnamefont{S.~M.} \bibnamefont{Ransom}},
  \bibinfo{author}{\bibfnamefont{M.}~\bibnamefont{Roberts}}, \bibnamefont{and}
  \bibinfo{author}{\bibfnamefont{J.~W.~T.} \bibnamefont{Hessels}},
  \bibinfo{journal}{Nature} \textbf{\bibinfo{volume}{467}},
  \bibinfo{pages}{1081} (\bibinfo{year}{2010}).

\bibitem[{\citenamefont{Alford et~al.}(2001{\natexlab{b}})\citenamefont{Alford,
  Bowers, and Rajagopal}}]{LOFF1}
\bibinfo{author}{\bibfnamefont{M.}~\bibnamefont{Alford}},
  \bibinfo{author}{\bibfnamefont{J.~A.} \bibnamefont{Bowers}},
  \bibnamefont{and}
  \bibinfo{author}{\bibfnamefont{K.}~\bibnamefont{Rajagopal}},
  \bibinfo{journal}{Phys. Rev. D} \textbf{\bibinfo{volume}{63}},
  \bibinfo{pages}{074016} (\bibinfo{year}{2001}{\natexlab{b}}).

\bibitem[{\citenamefont{Bowers and Rajagopal}(2002)}]{LOFF2}
\bibinfo{author}{\bibfnamefont{J.~A.} \bibnamefont{Bowers}} \bibnamefont{and}
  \bibinfo{author}{\bibfnamefont{K.}~\bibnamefont{Rajagopal}},
  \bibinfo{journal}{Phys. Rev. D} \textbf{\bibinfo{volume}{66}},
  \bibinfo{pages}{065002} (\bibinfo{year}{2002}).

\bibitem[{\citenamefont{Casalbuoni and Nardulli}(2004)}]{Casalbuoni04}
\bibinfo{author}{\bibfnamefont{R.}~\bibnamefont{Casalbuoni}} \bibnamefont{and}
  \bibinfo{author}{\bibfnamefont{G.}~\bibnamefont{Nardulli}},
  \bibinfo{journal}{Rev. Mod. Phys.} \textbf{\bibinfo{volume}{76}},
  \bibinfo{pages}{263} (\bibinfo{year}{2004}).

\bibitem[{\citenamefont{Paulucci et~al.}(2011)\citenamefont{Paulucci, Ferrer,
  de~la Incera, and Horvath}}]{Paulucci2011}
\bibinfo{author}{\bibfnamefont{L.}~\bibnamefont{Paulucci}},
  \bibinfo{author}{\bibfnamefont{E.~J.} \bibnamefont{Ferrer}},
  \bibinfo{author}{\bibfnamefont{V.}~\bibnamefont{de~la Incera}},
  \bibnamefont{and} \bibinfo{author}{\bibfnamefont{J.~E.}
  \bibnamefont{Horvath}}, \bibinfo{journal}{Phys. Rev.D}
  \textbf{\bibinfo{volume}{83}}, \bibinfo{pages}{043009}
  (\bibinfo{year}{2011}).

\bibitem[{\citenamefont{Alford et~al.}(2005)\citenamefont{Alford, Braby, Paris,
  and Reddy}}]{AlfordReddy}
\bibinfo{author}{\bibfnamefont{M.}~\bibnamefont{Alford}},
  \bibinfo{author}{\bibfnamefont{M.}~\bibnamefont{Braby}},
  \bibinfo{author}{\bibfnamefont{M.}~\bibnamefont{Paris}}, \bibnamefont{and}
  \bibinfo{author}{\bibfnamefont{S.}~\bibnamefont{Reddy}},
  \bibinfo{journal}{Astrophys. J.} \textbf{\bibinfo{volume}{629}},
  \bibinfo{pages}{968} (\bibinfo{year}{2005}).

\bibitem[{\citenamefont{Davies et~al.}(2010)}]{HPQCD}
\bibinfo{author}{\bibfnamefont{C.~T.~H.} \bibnamefont{Davies}}
  \bibnamefont{et~al.}, \bibinfo{journal}{Phys. Rev. Lett}
  \textbf{\bibinfo{volume}{104}}, \bibinfo{pages}{132003}
  (\bibinfo{year}{2010}).

\bibitem[{\citenamefont{Nakamura et~al.}(2010)}]{PDG}
\bibinfo{author}{\bibfnamefont{K.}~\bibnamefont{Nakamura}}
  \bibnamefont{et~al.}, \bibinfo{journal}{J. Phys. G}
  \textbf{\bibinfo{volume}{37}}, \bibinfo{pages}{075021}
  (\bibinfo{year}{2010}).

\bibitem[{\citenamefont{Benvenuto and Horvath}(1989)}]{BH89}
\bibinfo{author}{\bibfnamefont{O.~G.} \bibnamefont{Benvenuto}}
  \bibnamefont{and} \bibinfo{author}{\bibfnamefont{J.~E.}
  \bibnamefont{Horvath}}, \bibinfo{journal}{Mon. Not. R. astr. Soc.}
  \textbf{\bibinfo{volume}{241}}, \bibinfo{pages}{43} (\bibinfo{year}{1989}).

\bibitem[{\citenamefont{Steiner et~al.}(2010)\citenamefont{Steiner, Lattimer,
  and Brown}}]{Lattimer}
\bibinfo{author}{\bibfnamefont{A.~W.} \bibnamefont{Steiner}},
  \bibinfo{author}{\bibfnamefont{J.~M.} \bibnamefont{Lattimer}},
  \bibnamefont{and} \bibinfo{author}{\bibfnamefont{E.~F.} \bibnamefont{Brown}},
  \bibinfo{journal}{Astrophys. J.} \textbf{\bibinfo{volume}{722}},
  \bibinfo{pages}{33} (\bibinfo{year}{2010}).

\bibitem[{\citenamefont{Rawls et~al.}(2011)}]{Rawls}
\bibinfo{author}{\bibfnamefont{M.~L.} \bibnamefont{Rawls}}
  \bibnamefont{et~al.}, \bibinfo{journal}{Astrophys. J.}
  \textbf{\bibinfo{volume}{730}}, \bibinfo{pages}{25} (\bibinfo{year}{2011}).

\end{thebibliography}

\end{document}